\documentclass{article}

\usepackage{microtype}
\usepackage{graphicx}
\usepackage{subfigure}
\usepackage{booktabs} 
\usepackage{amsmath}
\usepackage{amsthm}
\usepackage{pgfplotstable}
\usepackage{pgfplots}
\usepackage{todonotes}
\pgfplotsset{compat=1.16}

\usepackage{hyperref}

\usepackage[accepted]{icml2020}

\icmltitlerunning{Scalable Econometrics on Big Data -- The Logistic Regression on Spark}

\begin{document}

\twocolumn[
\icmltitle{Scalable Econometrics on Big Data -- The Logistic Regression on Spark}

\icmlsetsymbol{equal}{*}

\begin{icmlauthorlist}
\icmlauthor{Aurélien Ouattara}{amzn}
\icmlauthor{Matthieu Bulté}{amzn}
\icmlauthor{Wan-Ju Lin}{tum}
\icmlauthor{Philipp Scholl}{tum}
\icmlauthor{Benedikt Veit}{tum}
\icmlauthor{Christos Ziakas}{tum}
\icmlauthor{Florian Felice}{amzn}
\icmlauthor{Julien Virlogeux}{amzn}
\icmlauthor{George Dikos}{amzn}
\end{icmlauthorlist}

\icmlaffiliation{amzn}{Amazon, Luxembourg}
\icmlaffiliation{tum}{Technical University of Munich, Germany}

\icmlcorrespondingauthor{Aurélien Ouattara}{amiyao@amazon.com}

\icmlkeywords{Big Data, Econometrics, Logistic Regression, Spark}

\vskip 0.3in
]



\printAffiliationsAndNotice{} 

\begin{abstract}
Extra-large datasets are becoming increasingly accessible, and computing tools designed to handle huge amount of data efficiently are democratizing rapidly. However, conventional statistical and econometric tools are still lacking fluency when dealing with such large datasets. 
This paper dives into econometrics on big datasets, specifically focusing on the logistic regression on Spark. We review the robustness of the functions available in Spark to fit logistic regression and introduce a package that we developed in PySpark which returns the statistical summary of the logistic regression, necessary for statistical inference. 
\end{abstract}
\section{Introduction}
\label{Introduction}

Extra-large datasets of billions of observations are becoming easily accessible with the popularization of computer and data based solutions and the soaring of technology companies \cite{varian}. Simultaneously, parallel computing methods (either through cluster based distributed computing or using the natural parallelization of Graphics Processing Units) are developing rapidly with the democratization of cloud-based on-demand solutions such as Amazon Web Services, Google Cloud or Microsoft Azure Cloud \cite{parallel}. By enabling economical and efficient solutions for handling large amount of data (especially by accelerating matrix-based computation), those tools are supporting the rapid development of machine learning methods \cite{varian_2018}. Parallel computing is also largely used to speed up data pre-processing (using SQL on Apache Spark for example), or exploratory data analyses (e.g. outlier detection, feature statistics or graphical analysis). 

Although private and public economic and financial studies would benefit from using more detailed and granular datasets \cite{data-revolution},  conventional statistical and econometric tools used for inference (such as Stata, R or even Python when it comes to larger datasets) are not able to support efficiently such models for big data\footnote{The phrase \textit{big data} usually entails two different concepts:  datasets with large number of observations or datasets with large number of features (usually equal to or larger than the number of observations.). In this paper, we will only refer to the former. A large litterature already specializes in the econometrics treatment of high-dimensionality datasets \cite{inference-high-dimensional}} with traditional machines. 

In this context, parallel computing is particularly attractive for economists and econometricians to analyze large amounts of data \cite{parallel}. However, with their main applications focused on Machine Learning methods, the newly developed tools optimized for handling large amount of data\footnote{Two of the most popular tools optimized for parallel computing, Spark and Tensorflow were developed in 2009 \cite{Spark} and in 2011 \cite{tensorflow2015-whitepaper} respectively.} usually support classical econometrics models and regressions only through their ‘machine learning' form and application, and therefore return error rates and information on the fit, but lack the outputs necessary for statistical inference usually found in the summary statistics (coefficient p-value, standard error, etc.).  

While sampling could prove a good alternative to the computational difficulties occurring with large data-sets \cite{wooldridge}, it requires additional verification on the sampling selection to avoid selection bias \footnote{Although large data-sets tend to allow for random samples representative of the population, one needs to apply particular care to avoid selection bias or omission of population categories (in the case of numerous categorical variables) in the selection mechanism.}. 

In this paper, we focus on the development of a robust econometric toolkit for logistic regression in PySpark, relying on the Machine learning library already built in PySpark: spark.ml \cite{spark_class_reg_doc}. 
Apache Spark is an open source framework optimized for the distributed computing of large datasets on computer clusters \cite{Spark}. A more detailed introduction to Spark in the context of economics studies can be found in Bluhm and Cutura (\citeyear{Bluhm}). 
The logistic regression is a standard binary response model\footnote{Binary response models are used when the dependent variable $y$ is binary, e.g. a variable returning whether a patient is infected}, of the form: 
\begin{equation}  \label{eq:model}
\mathbf{P}(y=1|X)=\frac{exp(X\beta)}{1+exp(X\beta)}
\end{equation}
With $y$ the dependent variable, $X$ the matrix of independent variables and $\beta$ the parameters. This model is particularly popular for econometric studies since it allows to easily calculate average partial effects for binary response models. It is also gaining in popularity as a standalone classification method or as an activation function in neural networks \cite{ML}. The logistic regression is usually solved by maximum likelihood estimations, using numerical optimization methods \cite{wooldridge}. Due to its popularity in machine learning, recent analytical tools integrate optimized solvers to fit logistic regressions, and because of the necessary numerical approximation, they oftentimes use different methods to find the parameters. All those reasons make logistic regression a particularly attractive model to tackle to initiate the development of econometric tools for big data. 

To develop a robust statistical toolkit for logistic regression, we first explore the current functions to fit logistic regression in Spark. Section \ref{sec:review} provides an overview of the difference between the Spark functions for logistic regression and the R method with respect to the consistency of the estimator returned. 
In Section \ref{sec:stat}, we present our computation of the statistical summary package (which includes a heteroskedasticity-robust standard error) for the logistic regression, specifically focusing on our implementation and optimization of the calculation of the covariance matrix in the Spark environment. 
Section \ref{sec:conc} concludes.

\section{Review of the Spark Functions for Logistic Regression}
\label{sec:review}

While looking into the different implementations of the logistic regression in Spark, we have observed a significant gap in the estimators returned by the different functions for the same dataset. 
We therefore start our analysis by exploring the difference in the estimators returned by different functions for the logistic regression, comparing the results with the R function \texttt{glm}, standard R function to fit Generalized Linear Models, including logistic regressions, referred to as \texttt{(GLM)}. Two implementations of the logistic regression are available in Spark (Spark ML library\footnote{Spark ML is the newer SciKit-Learn inspired machine learning library of Spark, which replaces Spark MLlib currently in maintenance/bug-fix only mode.}), the {\small \texttt{ml.regression.GeneralizedLinearRegression}}, which we will refer to as \texttt{(GLR)} and the \texttt{ml.classification.LogisticRegression} referred to as \texttt{(LR)}. In a first part, we dive into the empirical evidence of the gap between the different implementations, which we explain later on by the difference in the solver used by both functions.

\subsection{Empirical Exploration of Spark's Logistic Regression Functions} \label{sec:first-exp}

Considering the maturity of R and its prominent role in the statistical world, we decide to treat \texttt{GLM} as an empirical reference and run our comparative analysis against the results provided by this method.
The first step of our analysis is to empirically assess the quality of the estimates for the parameters $\beta$ present in (\ref{eq:model}). This assessment is performed on a sub-sample of an applied dataset which consists of 100 million of observations and 140 variables in total (with a majority of categorical variables one-hot encoded into multiple dummy variables).

\begin{figure*}[ht] 
\begin{center}
\begin{subfigure}
\centering
\includegraphics[width=\columnwidth]{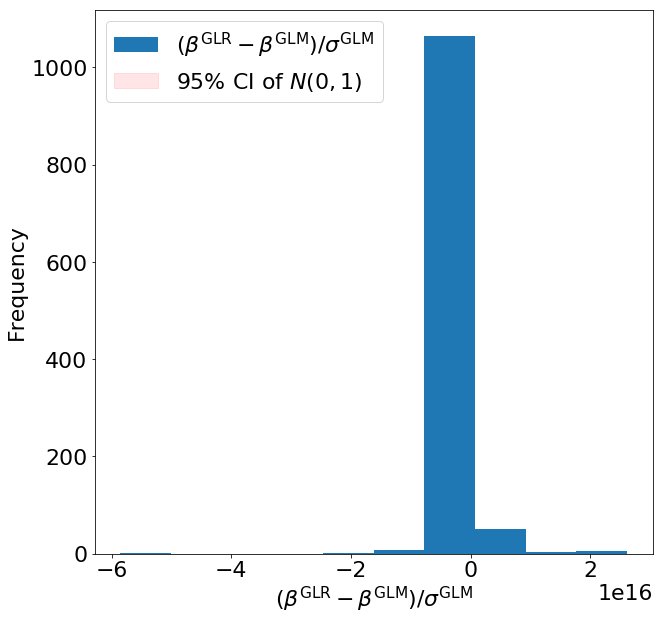}
\end{subfigure}%
\begin{subfigure}
\centering
\includegraphics[width=\columnwidth]{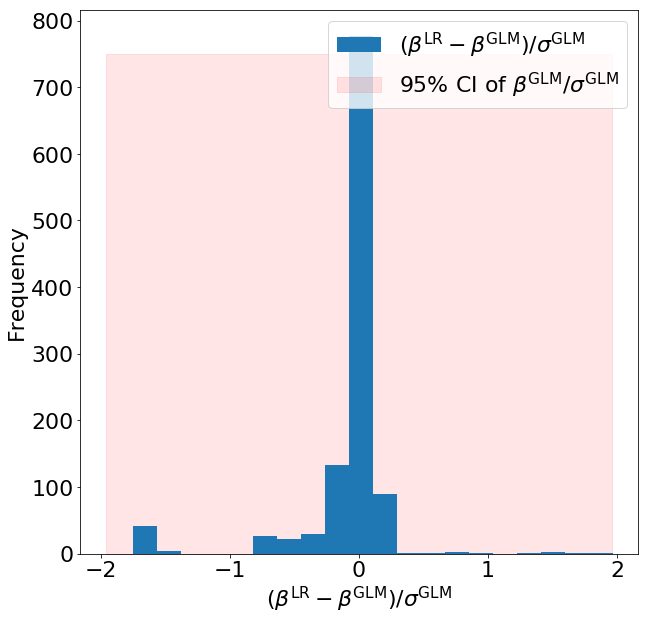}
\end{subfigure}
\caption{Empirical comparison of the coefficient estimators applied to 10 subsets of 100K observation. In each panel, we show the histogram of the variance-adjusted error of every coefficient. The top and down histograms are respectively based on the \texttt{GLR} and \texttt{LR} methods. The mean and standard deviation of the errors are respectively $(-3.56 \cdot 10^{13}, 2.91 \cdot 10^{15})$ and $(-9.24 \cdot 10^{-2}, 3.80 \cdot 10^{-1})$ for the \texttt{GLR} and \texttt{LR} methods. Note that the x-axis of the top panel has a scale of $10^{16}$.}
\label{fig:comparative-all-cols}
\end{center}
\end{figure*}

Let $\hat{\beta}^{glm}$, $\hat{\beta}^{glr}$ and $\hat{\beta}^{lr}$ denote the estimators computed by each of the considered methods. The aim of the empirical analysis is to estimate the standardized errors between the Spark estimators and the R estimators:
\begin{equation} \label{eq:err_spark}
    	\widehat{\text{err}}^{spark} = \frac{\hat{\beta}^{spark} - \hat{\beta}^{glm}}{\hat\sigma^{glm}}
\end{equation}
To that end, we randomly sample $10$ subsets of $100$K from our main dataset observations and return the three estimators on each subsample. This gives us an empirical estimation of the distribution of $\widehat{\text{err}}^{glr}$ and $\widehat{\text{err}}^{lr}$. In our dataset example, the \texttt{LR} method reports satisfactory results -- every coefficient lay within the theoretical 95\% confidence interval around the value computed by \texttt{GLM} -- whereas the \texttt{GLR} method proves to be unstable with standardized errors as high as $10^{16}$. By exploring our dataset further and by replicating the analysis after removing variables, we observe that this instability is driven by the presence of "quasi-multicollinearity" in our predictors -- representing the generic situation where the eigenvalues of $X^\top X$ are almost equal to $0$. As a result, the value of the determinant of $(X^\top X)^{-1}$ -- proportional to the variance of our coefficients -- explode. Interestingly in our case, the low eigenvalues are not driven by high variable to variable correlation coefficients, but by a high correlation of linear combinations of our variables (due to the high number of categorical variables in our initial dataset). This case is therefore not identifiable by simple correlation analysis of our dataset.
The results of this first benchmark are illustrated in Figure \ref{fig:comparative-all-cols}.

\subsection{Theoretical Comparison of the Implementations of the Logistic Regression} \label{sec:theoretical-results}

This section dives into the source code of the different methods tested above, and compare the construction and implementation of the three estimators considered in this analysis. All three estimators reviewed attempt to approximate the parameter vector $\beta$ by maximum likelihood estimation (MLE) \cite{r_src_glm, spark_adtop, spark_class_reg_doc} using second order methods: both \texttt{GLM} and \texttt{GLR} methods solve the optimization problem using a specialized Newton method, the Iteratively reweighted least squares (IRLS) algorithm \cite{r_src_glm,spark_src_glr}. The \texttt{LR} method approximates the MLE using a quasi-Newton method, the Limited-memory BFGS (L-BFGS) algorithm \cite{spark_src_lr}.  

The main difference between both solvers is that the IRLS requires a recalculation and inversion of the Hessian for each iteration. 
In addition to the computational price of inverting the hessian at each step, this poses a risk of numerical instability if the matrix is ill-conditioned. This explains why the \texttt{GLR} function in Spark is particularly sensitive to high correlation or quasi multi-collinearity in the dataset. In such a case, the \texttt{GLR} function attempts to invert an ill-conditioned matrix by using the Cholesky-Decomposition \cite{spark-glr-cholesky}, which leads to the instability observed. On the other hand, the \texttt{GLM} function uses the QR-decomposition with column pivoting to solve the weighted least square in each of the iteration of the IRLS algorithm in a lower dimensional subspace of the feature space -- therefore removing the low eigenvalue variables. The problem is then well-posed and better conditioned although it solves for a slightly modified version of the problem \cite{r-glm}.

In order to simplify the computation of Newton's method, Quasi-Newton methods use an approximation  of the Hessian matrix. The Limited Memory Broyden-Fletcher-Goldfarb-Shanno (L-BFGS) directly calculates an approximation of the invert of the Hessian, following the BFGS principles \cite{BFGS}. In addition to improving the memory usage, this method also presents the benefit of not inverting the matrix directly  \cite{L-BFGS}, which limits the potential instability driven by the inversion step. This explains why we see that the \texttt{LR} function in Spark is less sensitive to high level of correlation than the \texttt{GLR} function.

\subsection{Statistical Evidence of the Theoretical Difference in Algorithms} \label{sec:test_fram}
Building from the difference in implementation between all three methods and to validate the observations from part \ref{sec:first-exp}, we review statistically the robustness of the estimators using artificially generated datasets with known parameters. With a large enough dataset, we expect the estimator to be statistically significantly equal to the known parameter. 

We sample an artificial feature matrix $X$ and parameters $\beta$ from multivariate Gaussians using  pseudo-random number generators (PRNG\footnote{In order to sample a large amount of artificial data efficiently, we use the PCG-XSH-RR-32 function in Cython \cite{behnel2010cython}. Using Cython allows us to reduce the time required to generate 100GB of random data from 8 days with Python to 4 minutes. This function is based on Permuted Congruential Generators (PCGs) which apply Linear Congruential Generators (LCGs) \cite{handbook-compstat} and different permutations to increase the randomness of specific bits \cite{oneill:pcg2014}. This function allows for the parallelizability of the data generation without loss in quality and demonstrates a strong statistical quality by passing the test suites TestU01 \cite{TestU01} and PractRand \cite{practrand}.}) \cite{handbook-compstat}. The response variable is sampled by 
$y_i\sim Bern(\sigma(x_i^T\beta))$, with $\sigma(x)=\frac{1}{1+\exp(-x)}$ the sigmoid function. 
Using those generated datasets, we compute the estimators $\hat{\beta}$ of $\beta$ for each function and review statistically the difference between $\hat{\beta}$ and $\beta$. 
We first observe that \texttt{LR} and \texttt{GLR} return unbiased estimator for a feature matrix with independent columns and balanced outcome. We therefore generate 1,000 datasets with 10 columns and 1M rows with these properties. For both algorithms, we compute 1,000 p-values (one p-value per dataset), which we expect to follow a uniform distribution. A Kolmogorov-Smirnov test against the uniform distribution, yields p-values of 0.5547 and 0.5548 for \texttt{LR} and \texttt{GLR} respectively and validates this hypothesis statistically.  

\begin{figure}[h]
	\centering
	\begin{subfigure}
	\centering
		\includegraphics[width=\columnwidth]{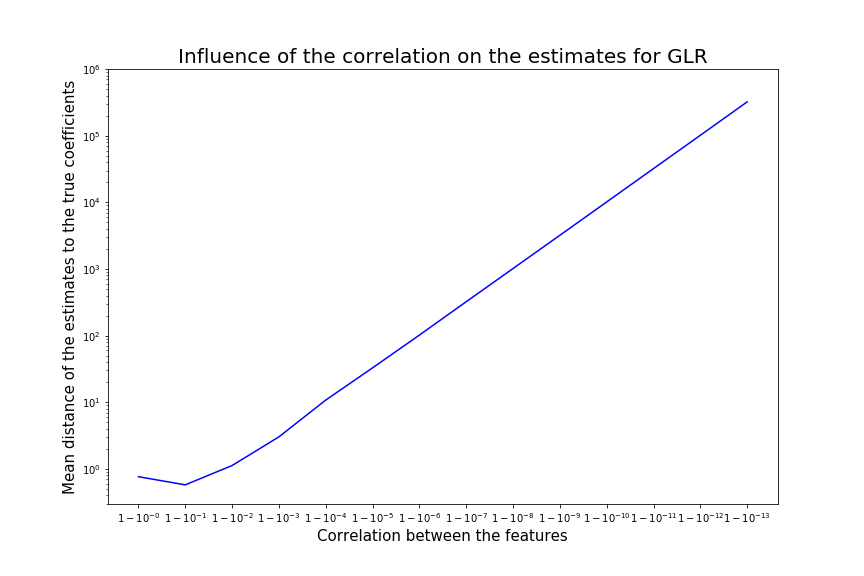}
		\end{subfigure}%
  	\begin{subfigure}
  	\centering
		\includegraphics[width=\columnwidth]{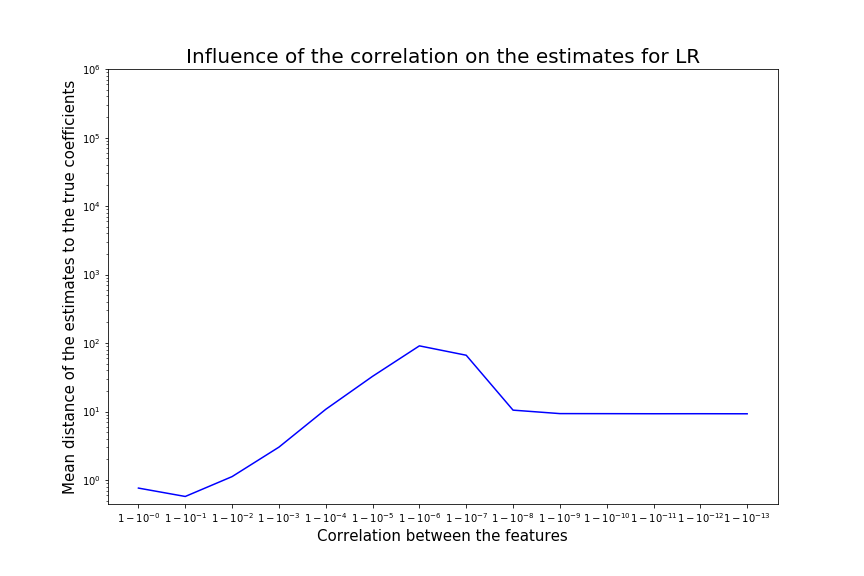}
	\end{subfigure}
 	\caption{Plots of the relation between the correlation between features and the mean distance between $\beta$ and $\hat{\beta}$. Left: \texttt{GLR}. Right: \texttt{LR}.}
  	\label{fig:lr-corr-data}
\end{figure}

We then review the robustness of both algorithms when confronted to highly correlated data, similar to our initial production dataset. For this, we test datasets with 1,000 rows and 2 correlated columns. We present the evolution of the results as a function of the correlation level ($1-0.1^k$ for $k=1,...,14$) between the two features in Figure \ref{fig:lr-corr-data}. We see that both algorithms perform similarly when the correlation is controlled (i.e. $<0.99$). However, we observe that \texttt{GLR}'s error increases as the correlation increases, while the results of \texttt{LR} remain more controlled. This is in adequation with the algorithmic difference assessed between both functions in part \ref{sec:theoretical-results}. As a result, we propose to use the \texttt{LR} function in the context of statistical studies.

\section{Computation of the Statistical Package} \label{sec:stat}
In addition to the pure fit of the model and the consistency of the results, we are also interested in the statistical significance of the model and its estimators to draw statistical inference. Since, \texttt{LR}, the robust function for logistic regression in Spark, as presented in section \ref{sec:theoretical-results} does not return a statistical summary, we propose in this section an implementation of a statistical summary for logistic regressions in Spark which resembles the summary of the R-package \texttt{GLM} \cite{glm}, even extending it by adding the heteroskedasticity robust sandwich errors estimates \cite{carroll1998sandwich}.

\subsection{Computation of the Statistical Summary} \label{sec:cov_mat}
The main challenge for the computation of the statistical summary is to develop a scalable implementation of the asymptotic variance matrix. Since the \texttt{LR} function fits the maximum-likelihood, the estimator of the asymptotic covariance matrix is the inverse of the Hessian matrix \cite{wooldridge}:
\begin{equation} \label{eq:MLE}
    \widehat{Avar(\hat{\beta})}=[X^TD(\hat{\beta})X]^{-1}
\end{equation}
with feature matrix $X$ and $D(\hat{\beta})$ a diagonal matrix with elements $d_i=\hat{p}(x_i)(1-\hat{p}(x_i))$ where $\hat{p}(x_i)=\frac{1}{1+\mathrm{e}^{-x_i\hat{\beta}}}$ is 
the predicted probability for the i-th observation.

This formulation of the covariance matrix assumes that the residuals of the underlying linear equation of the logistic regression have homogeneous variance. In general, this is often not true, especially when dealing with categorical exogenous variables. Therefore, we present the sandwich error to our summary as it is robust to heteroskedasticity. The formula for this error is given by replacing the estimator of the asymptotic covariance matrix by \cite{carroll1998sandwich} 
\begin{equation}
    \hat{V}=A_n^{-1}(\hat{\beta})B_n(\hat{\beta})A_n^{-1}(\hat{\beta})
\end{equation}
with $A_n$ the MLE covariance matrix from (\ref{eq:MLE}) divided by the number of observations, $n$ and 
\[B_n=n^{-1}\sum_{i=1}^n x_ix_i^T (Y_i-p(x_i))^2/(1-h_{ii}),
\] where 
\[h_{ii}=p(x_i)(1-p(x_i))x_i^TA_n^{-1}(\hat{\beta})x_i\] correspond to the diagonal entries of the hat matrix.

From the covariance matrix (or Sandwich covariance matrix), we derive the main statistics such as the standard errors, t-statistics, p-values and confidence intervals. We also return performance statistics such as the Akaike information criterion (AIC), McFadden's $\mathrm{Pseudo}$-$R^2$ and the likelihood-ratio test, which are derived from the log-likelihood of the model, as well as the Receiver Operating Characteristic (ROC) curve to complete this set of statistics. 

To validate the robustness of our implementation of the covariance matrix empirically, we compute the empirical covariance matrix using bootstrap for comparison with our variance expressed with the theretical function. The bootstrap estimation of the variance consists in a replication a certain number of time of the maximum likelihood estimation (100, 1000, 10000) and the computation of the variance of the betas for each batch generated. Similar to the review the estimators using artificial data as presented above, this methodology allows us to validate the robustness of our methodology to compute the covariance matrix.

It is important to note that using the p-value as a measure for statistical significance requires adjustments in the treatment of large data-sets as presented in \cite{large_sample_p_value}. Our methodology returns the p-value without extrapolation on the significance level and we therefore let the interpretation of the value to the user. 

\subsection{Scaling up the Generation of the Statistical Summary}

In order to propose scalable statistics on large datasets, we aim to reduce the computation time and/or computing cost by ensuring that all computing actions are optimized. We present in this part the steps that we have taken to optimize our code using Spark native functionalities. 

As reviewed in part \ref{sec:cov_mat}, our implementation is relying on linear algebra computations. To optimize our performance on linear algebra calculations, we compare BLAS \cite{blas_docu} called directly with \textit{scipy.linalg} and NumPy \cite{numpy_docu}, which are two standard packages in Python supporting linear algebra. A comparison on matrix addition\footnote{\texttt{daxpy} in BLAS and \texttt{add} in NumPy}, dot product\footnote{\texttt{ddot} in BLAS and \texttt{dot} in NumPy} and outer dot product\footnote{\texttt{dgem} in BLAS and \texttt{outer} in NumPy} computations between both packages show that BLAS is three times faster than NumPy on our test sample. We therefore rely on BLAS for our matrix computations in our class function. 
We also use \texttt{mapPartitions} transformation\footnote{\texttt{mapPartitions} transformation applies a function to a partition of rows instead of each row independently and thus leads to a significant improvement in terms of memory usage and time.} to compute the inverse of the covariance matrix and the sandwich error within the same Spark operations. 

In addition, we also optimize our code using the mathematical properties of our matrices. For the computation of the standard covariance matrix, we first rewrite the Hessian as a Gramian matrix:
\begin{equation}
	X^TD(\hat{\beta})X= (D(\hat{\beta})^{\frac{1}{2}}X)^T(D(\hat{\beta})^{\frac{1}{2}}X)
\end{equation}
This allows us to rewrite the matrix multiplication as the sum of outer products of the columns in the first matrix with the corresponding rows in the second matrix:
\begin{equation}
    X^TD(\hat{\beta})X= \sum_{i=1}^n d_ix_ix_i^T
\end{equation}
This presents the advantage that less memory accesses are needed for the calculations, reducing the computation time of the covariance matrix from 2.5h to about 15mins. 

For the Sandwich covariance matrix, we face an additional problem when computing the elements $h_{ii}$ as we have to perform a matrix-vector and a vector-vector-multiplication for each element. In order to improve the performance and numerical stability, we calculate $A_n^{-1}x$ by first computing the Cholesky decomposition of $A_n$ and then use forward/back substitution to solve for $A_n^{-1}x$. This approach reduces the computation time of the sandwich error by half on our test data sample.

\begin{figure}[ht]
	\centering
	\includegraphics[width=\columnwidth]{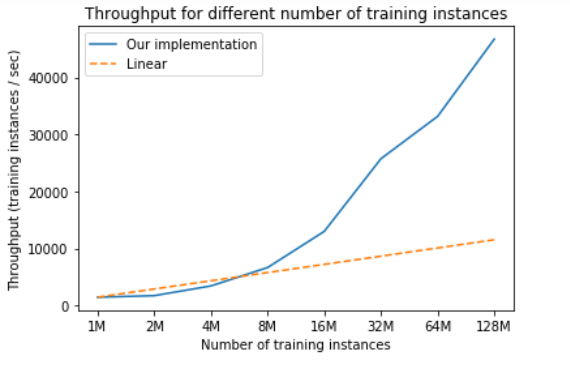}
 	\caption{Throughput (training instances / sec) of the implementation of the statistical summary for different number of training instances.}
  	\label{fig:throughput}
\end{figure}

In order to evaluate the scalability of our implementation of the statistical summary, we measure the throughput, which is the number of processing training instances per second, for 1, 2, 4, 8, 16, 32, 64 and 128 millions of training instances. The experiment runs on an Amazon Web Services EMR cluster of 1 m4.2xlarge primary node and 10 c4.4xlarge worker nodes, through a Apache Zeppelin notebook. The results of our experiment, illustrated in Figure \ref{fig:throughput}, show that our implementation can efficiently be used for statistical analysis on vast amounts of data. Indeed, the throughput in our implementation increases almost exponentially when the size of the dataset increases, thus validating the scalability of our implementation.

\subsection{Presentation of the Statistical Package in PySpark} 
\begin{figure*}[h]
	\centering
	\includegraphics[width=15cm]{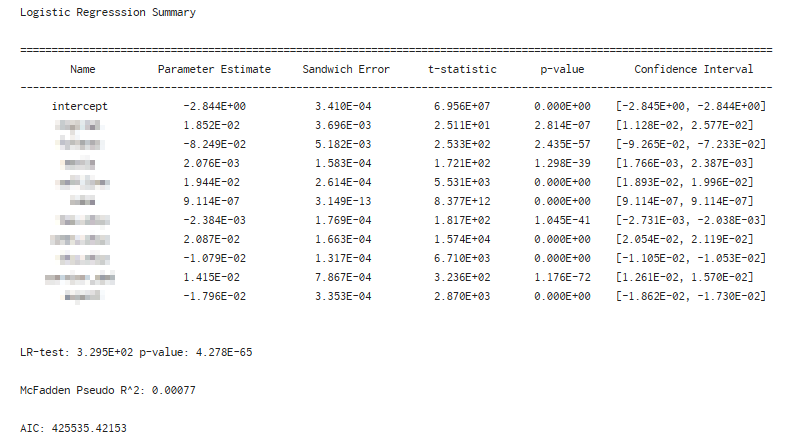}
 	\caption{Format of the summary output}
  	\label{fig:summary_output}
\end{figure*}

In order to deliver a shareable and re-usable implementation, all methods and variables required for the summary are included in a class. Users are able to specify whether the intercept is considered in the analysis or whether the standard error is returned -- the sandwich error being used by default. The class method \texttt{fit} computes all variables sequentially and returns the summary in a similar format to the \texttt{glm} function in R. The arguments that \texttt{fit} method takes are the dataset, the names of each explanatory variable, the name of the response, and the boolean variables corresponding to the user’s specifications. The dataset is a Spark \texttt{DataFrame} consisting of two columns, the first \texttt{Double} column contains the response and the second one contains all the explanatory variables in a sparse vector. The class returns the variables name, the estimators from \texttt{LR} and the variables presented above and is printed in a format similar to the \texttt{glm} function in R (see Figure \ref{fig:summary_output}).

\section{Conclusion and future works} \label{sec:conc}

In this paper we have highlighted the differences between the two functions to fit logistic regressions in Spark ML library. Based on our analysis, we recommend econometricians to use the \texttt{LR} function in order to get robust results, even under high correlation, in a similar fashion to the ones obtained with classical tools. Indeed, although data pre-processing and feature selection are commonly used in machine learning methods to deal with highly correlated variables, econometricians might decide to keep highly correlated (although non-colinear) variables to avoid model selection mistakes to draw inference, as highlighted in Belloni et al. (\citeyear{double-lasso}). In order to complete the use case for econometric study and to allow for drawing statistical inference with Spark's logistic regression, we have also developed a scalable software package based on Apache Spark to support statistical inference based on logistic regression by providing a statistical summary. This class presents a first attempt to propose scalable statistics for logistic regression on Spark, but its efficiency could be further improved, notably through a development in Scala, which is more efficient as native to Spark. We present in table \ref{tab:pareto_vs_mllib} a comparison\footnote{The comparison is made on an applied dataset of 10 covariates, using an Amazon Web Services EMR cluster of 1 m4.2xlarge primary node and 10 c4.4xlarge worker nodes} of the runtime of our function (fit and statistical summary) and of the \texttt{GLR} function. Although we do not recommend using \texttt{GLR} as described above, we use it for comparison purposes as this is the only function in Apache Spark which returns a statistical summary.

\begin{table}[!ht]
    \centering
    \begin{tabular}{rcc}
        \hline\hline
        $n$ &  Our Implementation & GLR \\ \hline
        1M & 0:00:27 & 0:00:10 \\
        5M & 0:01:08 & 0:00:31 \\
        10M & 0:02:42 & 0:00:52 \\
        50M & 0:24:22 & 0:29:33 \\
        100M & 0:56:53 & 1:26:51 \\
        \hline
        Summary & Yes & Yes \\
        Robust s.e. & Yes & No \\
        \hline\hline
    \end{tabular}
    \caption{Run time comparison between GLR and our implementation}
    \label{tab:pareto_vs_mllib}
\end{table}

Through our exploration of Apache Spark, we have experienced issues on expressiveness \footnote{Expressiveness is a qualitative property representing how well a concept, in our case mathematical, can be expressed within a programming language or computational framework.}, and performance. Indeed, although a strong framework for large scale distributed data cleaning and processing, the level of expressiveness of the Spark API makes it difficult to implement low level functionalities without sacrificing either code simplicity or performance. For this reason, in the context of building scalable statistical toolkits for economists and econometricians, it is also worth exploring alternatives to Spark for large scale high-performance computing such as TensorFlow and Dask. 

TensorFlow \cite{tensorflow} is a machine learning system developed by Google that operates at large scale and in heterogeneous environments. It is a well-known tool for deep learning where it utilizes dataflow graphs to represent the computations, shared states, and the operations mutating those states. Tensorflow optimizes the logistic regression using First Order Methods (Derivatives of Stochastic Gradient Descent) and uses GPUs for parallel computing. Both those properties show significant difference with the functions of Spark that we have reviewed in this document and therefore TensorFlow is of particular interest for a further exploration of statistical toolkit for logistic regression on large dataset. 

\section*{Acknowledgements}

Most of the work presented in this paper has been done as part of the Summer Semester 2019 Data Innovation Lab held by Technical university of Munich and Amazon. An earlier version of this work is available at \cite{TUM}. 
We thank Konstantinos Petropoulos and Martina Malfitana  for their  assistance  in  data  collection,   analysis and code review. We thank Victor Chernozhukov, Han Hong, Suresh Saggar and Ricardo Acevedo Cabra  for  providing insightful feedback, comments and reviews. We specifically thank Edouard Oddo for supporting this project.

\nocite{Spark_book,spark_overview,spark-glr-irls,pyspark_arch,shuffling,data_locality,garbage_collector,parquet,armbrust2015spark,urminskyusing,square_root_lasso,glm}
\bibliography{biblio}
\bibliographystyle{icml2020}

\end{document}